\documentclass[pdflatex,sn-mathphys-num]{sn-jnl}

\usepackage{siunitx}
\usepackage{graphicx}%
\usepackage{multirow}%
\usepackage{amsmath,amssymb,amsfonts}%
\usepackage{amsthm}%
\usepackage{mathrsfs}%
\usepackage[title]{appendix}%
\usepackage{xcolor}%
\usepackage{textcomp}%
\usepackage{manyfoot}%
\usepackage{booktabs}%
\usepackage{algorithm}%
\usepackage{algorithmicx}%
\usepackage{algpseudocode}%
\usepackage{listings}%
\usepackage{fix-cm}
\usepackage{mathrsfs}

\theoremstyle{thmstyleone}%
\theoremstyle{thmstyletwo}%

\theoremstyle{thmstylethree}%
\begin{document}
\title[Article Title]{\textbf{Ultra-Thin Aluminum-Doped Silver for Transmissive Thermally Reconfigurable Visible Photonics}}


\author[1,2]{\fnm{Hongyi} \sur{Sun}}
\author[1,2]{\fnm{Yi-Siou} \sur{Huang}}
\author[3]{\fnm{Junyeob} \sur{Song}}
\author[4]{\fnm{Francis} \sur{Vásquez-Aza}}
\author[5]{\fnm{Christopher S.}\sur{Whittington}}
\author[6]{\fnm{Nathan}\sur{Youngblood}}
\author[5,7]{\fnm{Sharon M.}\sur{Weiss}}
\author[4]{\fnm{Georges} \sur{Pavlidis}}
\author[8]{\fnm{Amit} \sur{Agrawal}}
\author*[1,2]{\fnm{Carlos A} \sur{Rios Ocampo}} \email{riosc@umd.edu}

\affil[1]{\orgdiv{Department of Materials Science and Engineering}, \orgname{{University of Maryland}, \city{College Park}, \country{USA}}}

\affil[2]{\orgdiv{Institute for Research in Electronics and Applied Physics}, \orgname{{University of Maryland}, \city{College Park}, \country{USA}}}

\affil[3]{\orgname{{National Institute of Standards and Technology}, \city{Gaithersburg}, \country{USA}}}
\affil[4]{ \orgdiv{Department of Mechanical Engineering},\orgname{University of Connecticut}, \city{Storrs}, \country{USA}}

\affil[5]{\orgdiv{Interdisciplinary Materials Science Program}, \orgname{{Vanderbilt University}, \city{Nashville}, \country{USA}}}

\affil[6]{\orgdiv{Department of Electrical Engineering},  \orgname{{University of Pittsburgh}, \city{Pittsburgh}, \country{USA}}}
\affil[7]{\orgdiv{Department of Electrical and Computer Engineering},  \orgname{{Vanderbilt University}, \city{Nashville}, \country{USA}}}

\affil[8]{\orgdiv{Department of Engineering}, \orgname{{University of Cambridge}, \city{Cambridge}, \postcode{CB3 0FA}, \country{UK}}}
\vspace{-1 cm}

\abstract{
Functional materials with high electrical conductivity and optical transmittance are vital for thermally tunable free-space photonic systems. Conventional transparent conductors such as graphene and indium tin oxide (ITO) are limited by high contact resistance, poor mechanical stability, or complex fabrication. Ultra-thin metals, such as pure silver, have also been explored with limited success due to thermal instability and dewetting. Here, we propose an ultra-thin aluminum-doped silver (Al-doped Ag) film to tackle these challenges.  Aluminum promotes heterogeneous nucleation of silver, enabling the formation of continuous, smooth films that are thermally stable at reduced thicknesses while maintaining excellent electrical conductivity and transparency. We find that a 12~nm Al-doped Ag film exhibits an average transmittance of \mbox{$\sim$80\%} across the visible range with a sheet resistance of {\unboldmath $8.3 \pm 1.16~\Omega/\mathrm{cm}^2$}. Moreover, on-chip Al-doped Ag microheaters exhibit uniform, rapid thermal response, and stable electrical performance, maintaining functionality for over {\unboldmath $10^7$} ON and OFF cycles at temperatures below 400~$^\circ$C. Furthermore, as a benchmark, we demonstrate reversible phase-change switching in Ge$_2$Sb$_2$Se$_4$Te (GSST) and VO$_2$. {\unboldmath 30$\times$30~$\mu$m$^2$} GSST cells exhibited complete crystallization and amorphization under 2.2~V–200~ms and 4.1~V–{\unboldmath 50~$\mu$s} pulses, respectively, resulting in a 40\% transmission contrast at 780~nm and a tenfold improvement in power consumption compared to similar devices. Additionally, VO$_2$ films displayed reversible insulator-to-metal transitions near 65~$^\circ$C with reflectance and transmittance modulation in the visible and the near-infrared at frequencies up to 25~Hz, with room for improvement. These results establish Al-doped Ag as a robust transparent metallic heater for integration in dynamic metasurfaces, optical coatings, and other reconfigurable photonic platforms.
}

\keywords{Optical materials, ultra-thin metals, phase-change materials}

\maketitle
\section{Introduction}\label{sec1}

Tunable materials that undergo refractive index modulation via the thermo‑optic effect \cite{liu2022thermo} or through heat‑triggered reversible phase transitions in phase-change materials (PCM) \cite{zhang2021electrically, king2024electrically, popescu2024electrically} are integral components of a wide range of advanced optical devices. Benefiting from a large index contrast, such devices include dynamic transmissive color filters\cite{he2020dynamically,huang2023tunable,xu2022passive}, reconfigurable metasurfaces capable of beam steering and image processing\cite{cotrufo2024reconfigurable,li2024phase}, and smart windows for infrared regulation\cite{jing2024long,yang2023phase}. To integrate such devices into free-space optical systems, an external heater is typically required to enable precise and large-area thermal control. \cite{papanastasiou2020transparent} In the visible spectral range, similar heaters have been demonstrated using materials such as graphene or indium tin oxide (ITO). However, both suffer from shortcomings. Graphene has high contact resistance and low fabrication yield \cite{rios2021multi}, while ITO has poor chemical stability and mechanical flexibility, which limits its compatibility with flexible substrates \cite{cairns2000strain}.  Moreover, ITO requires thick layers ($>$ 100 nm) and post-deposition high-temperature annealing for good conductivity \cite{kim2006thickness}.\

In contrast, thin metal heaters offer superior mechanical and electrical performance with consistent $>90$\% fabrication yields. However, a fundamental challenge arises from the inherent trade-off between optical transparency and electrical conductivity in thin-film metals: reducing film thickness improves transparency by allowing greater light transmission, but increased electron scattering at interfaces and grain boundaries increases electrical resistivity \cite{fahland2001low}. Moreover, the fabrication process significantly impacts the electrical performance of thin-film materials. In physical vapor deposition (PVD), the metal film grows on the surface as isolated islands before coalescing into a continuous film, due to its low adhesion energy to the substrates \cite{zheng2022grain}.  Therefore, a thin metal heater must exceed the critical thickness to achieve a continuous structure and reasonable conductivity. Furthermore, it is crucial to suppress the rough surface morphology of thin continuous metal films, especially when heaters must operate at high temperatures \cite{alarcon2023ostwald,kraft2001fatigue}. \

Silver, the most electrically conductive metal in the periodic table, exhibits the lowest imaginary part of dielectric permittivity $\varepsilon_2$, among all metals in the visible and near-infrared spectral regions. This combination of low optical loss and high electrical conductivity makes silver a leading candidate for ultra-thin, transmissive films that maintain low resistance while maintaining high optical transparency \cite{zhang2021thin}. Moreover, the inherent ductility of metals means that ultra-thin silver films are excellent candidates for flexible transparent conductors \cite{ma2024pushing}. Nevertheless, depositing continuous, smooth ultra-thin silver films remains challenging due to poor wettability and island growth on dielectric substrates. A widely adopted approach to address these limitations employs an ultra-thin wetting layer to enhance adhesion between the silver heater and the transparent substrate. Transparent dielectric materials such as Al-doped ZnO, MgO, and TiO$_2$ can provide moderate wettability, thus improving film continuity and surface smoothness \cite{cho2011surface_alzno, lee2016mg, mosquera2016effect_TiO2}. Alternatively, metallic wetting layers have also been explored and generally yield smoother silver films than their dielectric counterparts \cite{chen2010ultra_Gewetting, wrobel2015ge, ciesielski2017controlling_NiGewetting, ji2025establishing_Ti}. However, metallic wetting layers tend to diffuse into silver over time, leading to the formation of segregated alloys. Since these wetting metals exhibit higher optical losses and electrical resistivity than silver, such interdiffusion inevitably degrades the optical and electrical performance of transparent heaters during operation \cite{ciesielski2018growth, todeschini2017influence}.\

An alternative strategy to achieve smooth ultra-thin silver films without an additional wetting layer involves doping a secondary metal directly into the silver matrix. In this approach, co-deposited additive metal atoms act as heterogeneous nucleation and clustering centers, altering the growth kinetics of silver and significantly reducing the percolation threshold on glass substrates \cite{zhang2017high}. This method has been applied to the optimization of silver thin films, where the dopant is introduced at low concentrations to preserve high optical transparency. For example, Zhang \emph{et al.} demonstrated a 7~nm Al-doped silver film with an average visible transmittance exceeding 80\% relative to a bare glass substrate \cite{zhang2014ultrathin, zhang2017high}. The aluminum dopants preferentially bond to oxygen at the surface of the substrate, providing stable heterogeneous nucleation sites for silver grains. Similarly, Ji \emph{et al.} reported an Al$_2$O$_3$/Cu-doped Ag/ZnO transparent electrode exhibiting a sheet resistance of 18.6~$\Omega/\mathrm{sq}$ and an absolute transmittance of 88.4\% \cite{ji2020ultrathin}.\

In this work, we exploit the high electrical conductivity and transparency of Al-doped Ag to experimentally demonstrate an ultra-thin transmissive microheater for tunable free-space optics in visible and near-IR wavelengths. To do so, we allow electrical current to flow through the Al-doped Ag thin-film, which creates a hotspot at the most resistive features via Joule heating, e.g., at the center of a bow-tie geometry. The devices exhibit excellent performance in electrical conductivity and optical transparency via optimized fabrication using a single-step co-sputtering and low-temperature post-annealing. We benchmark Al-doped Ag by reversibly switching volatile and nonvolatile PCMs in reconfigurable free-space optical devices.

\section{Results and discussion}\label{sec2}
\subsection{Al-doped Ag films: electrical and optical characterization}\label{subsec2_1}
\begin{figure}[h]
\centering
\includegraphics[width=\textwidth]{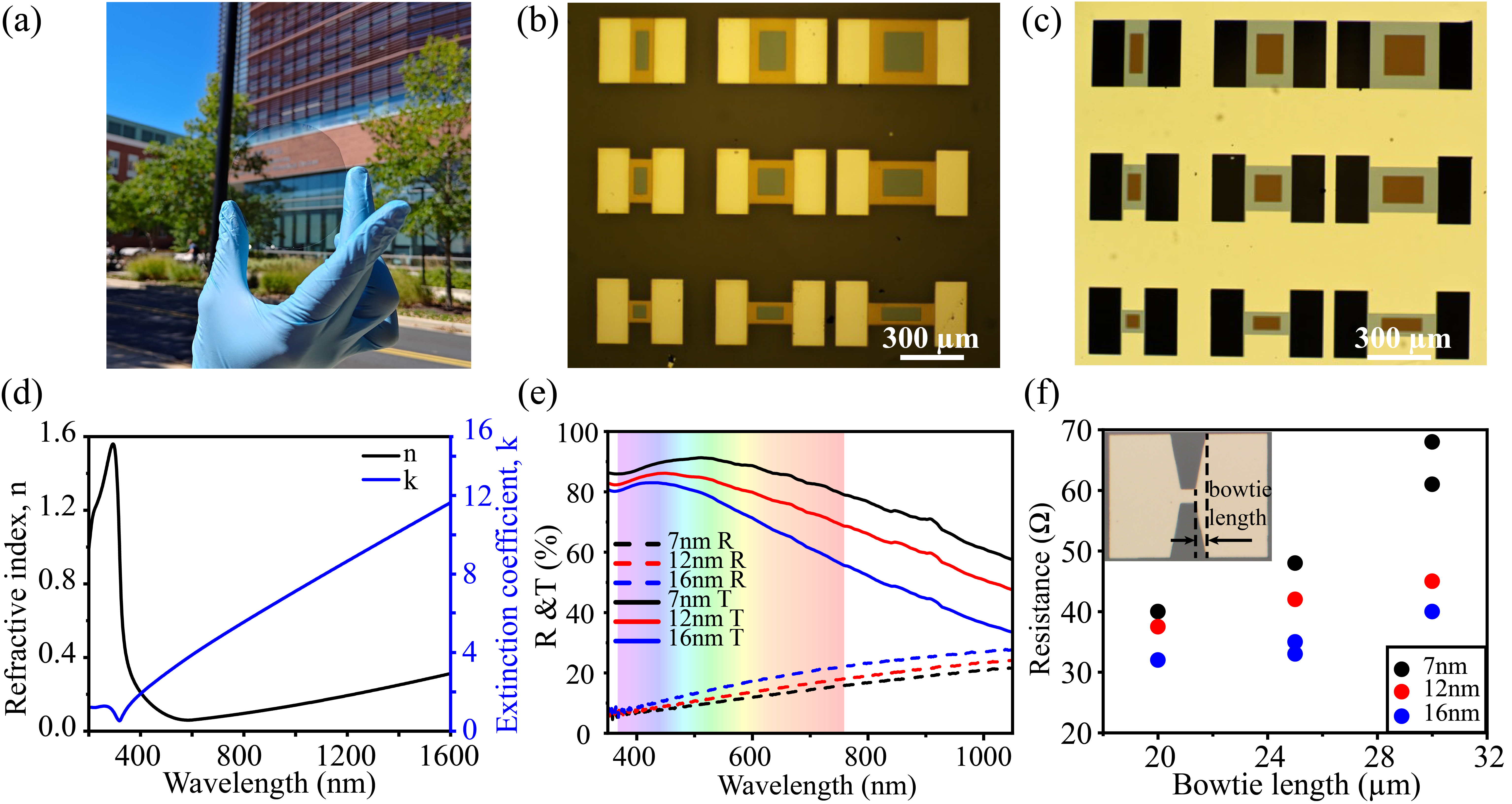}
\caption{Ultra-thin Al-doped Ag microheaters. (a) Optical image of the 12~nm-thick Al-doped Ag film over a 3~in sapphire wafer.(b) The reflection and (c) transmission images of multiple Al-doped Ag heaters with Ge$_2$Sb$_2$Se$_4$Te cells under an optical microscope with halogen illumination. (d) The refractive index and extinction coefficient from ellipsometry, measured on a 12~nm-thick Al-doped Ag film. (e) Reflectance and transmittance spectra in the visible and near-infrared for 7, 12, and 16~nm-thick Al-doped Ag thin films. (f) Electrical resistance measured on a 35$\times$35 $\mu$m$^2$ bowtie microheater with 7, 12, and 16~nm-thick Al-doped Ag films. }
\label{fig1}
\end{figure}

The Al-doped Ag-based microheaters were fabricated on 1 mm-thick fused silica and sapphire wafers. 7 nm, 12 nm, and 16 nm-thick Al-doped Ag films were deposited via a co-sputtering process, as described in detail in Methods and Materials, and following the method presented in \cite{zhang2017high}. Figure~\ref{fig1}(a) shows a photograph of a 10 nm-thick Al-doped Ag film on a 3-inch sapphire wafer, demonstrating the high transmission of this material platform. At the microscale and device level, Figures~\ref{fig1}(b) \& (c) show the optical images (under halogen illumination) of Al-doped Ag films after patterning via lithography and etching (see Methods and Materials) to form square microheaters. The devices are completed with Al electrodes and Ge$_2$Sb$_2$Se$_4$Te cells, demonstrating their versatility for both reflection and transmission measurements. We measured the refractive index and the extinction coefficient using ellipsometry in the 12 nm-thick sample, as shown in Figure~\ref{fig1}(d). The plasma frequency of the Al-doped Ag film was measured at 315 nm, indicating a minimal dopant-induced blue shift compared to the 320 nm plasma frequency of pure silver. The refractive index of Al-doped Ag is comparable to that of pure Ag, consistent with previous reports, while the extinction coefficient is slightly reduced. This indicates that the Al doping concentration is well optimized: neither too high to introduce defect-related losses nor too low to effectively improve the smoothness of the film \cite{zhang2014ultrathin}.\

Figure~\ref{fig1}(e) shows the optical transmission for 7, 12, and 16~nm thick Al-doped Ag films. The 7~nm-thick sample exhibits more than 80\% transmittance across the visible spectrum, with a peak of transmission of 90\% at 508~nm. The optical transmission decreased for the 12~nm- and 16~nm-thick films due to increased thickness, as the peak shifted to the blue. In contrast, the sheet resistance of the 7~nm-thick Al-doped Ag film was $18.7 \pm 0.54~\Omega/\mathrm{sq}$, decreasing to $8.3 \pm 1.16~\Omega/\mathrm{sq}$ and $7.0 \pm 1.04~\Omega/\mathrm{sq}$ for the 12~nm-thick and 16~nm-thick Al-doped Ag film. We further evaluated the electrical performance of the heater, as shown in Figure~\ref{fig1}(f), using 35~$\times$~35~$\mu$m$^2$ microheaters with three different bowtie lengths---defining the bowtie taper angle and the total length. As expected, the electrical resistance of the Al-doped Ag heater increases for smaller thicknesses and longer bowties, where the 12~nm thick devices demonstrated approximately 40~$\Omega$ for 20 $\mu$m, the shortest bowtie length tested. We note that while a shorter bowtie shows lower resistance, the limit where there is no tapering (i.e., a bowtie length of 0 ~$\mu$m, like those in Fig.~\ref{fig1}(b) \& (c) ) introduces unwanted current densities at the sharp corners, which can in turn dissipate heat and be less efficient at the center of the microheater.\

\subsection{Thermoreflectance characterization via \texorpdfstring{Ge$_2$Sb$_2$Se$_4$Te}{GSST} thermo-optical response}\label{subsec2_2}

\begin{figure}[!h]
\centering
\includegraphics[width=1\textwidth]{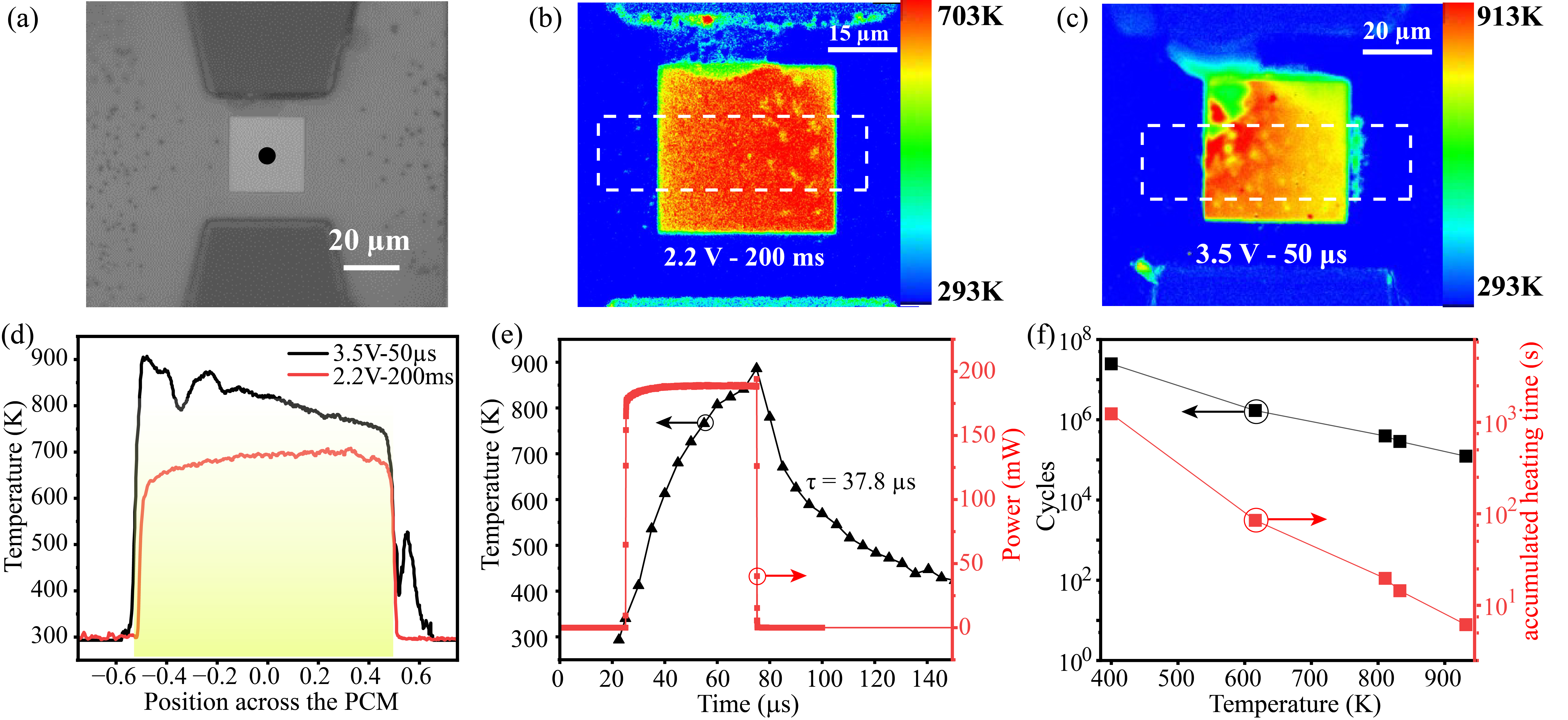}
\caption{\textbf{Thermal performance of Al-doped Ag heater using GSST thermo-optical response.} (a) Optical microscope image of a measured 25~$\times$~25~$\mu$m$^2$ heater with 30 nm-thick GSST.  Transient thermoreflectance images at the end of (b) 2.2 V - 200 ms and (c) 3.5 V - 50 $\mu$s pulses. The dashed lines delimit the area considered in the temperature profiles in (d). (d) Average temperature vs. position along the horizontal direction of the microheater bridge. (e) Experimental electrical pulse and real-time temperature at the center of the GSST pattern, marked in (a). (f) Endurance test: number of thermal cycles (black) and the accumulated heating time (red) vs. temperature at the end of a 50~$\mu$s-width pulse } \label{fig3}
\end{figure}

The transparency of Al-doped Ag films makes it difficult to use characterization techniques such as transient thermoreflectance imaging (TTI), which is widely used to study the transient and steady-state response of similar microheaters \cite{sun2025microheater}. To circumvent this challenge, we used a thin-film patch of crystalline Ge$_2$Sb$_2$Se$_4$Te (GSST) in close contact with Al-doped Ag films, as shown in Fig.~\ref{fig3}(a). We deposited an additional AlN layer between the microheater and the GSST to protect the Al-doped Ag film during lithography and patterning. GSST enables indirect characterization of the thermal dynamics of the microheater via its thermo-optical response, following the method presented in \cite{nobile2023time}. Fig.~\ref{fig3}(b)-(c) show the transient thermoreflectance images for two distinct scenarios: a mid-range temperature for an extended annealing time using 2.2 V - 200 ms on a 45$\times$45 $\mu$m$^2$ microheater, and a high-temperature for a time approaching the thermal time constants of the device with 3.5 V - 50 $\mu$s pulses on a 35$\times$35 $\mu$m$^2$ microheater. The former achieves a nearly uniform temperature profile across the GSST cell, with a 410 K rise from room temperature. The latter, on the other hand, shows non-uniform heating, with a maximum 900 K (nearly 610 K rise) and an 180 K temperature gradient across the GSST on a 45$\times$45 $\mu$m$^2$ microheater, which can be attributed to the heater not yet reaching the steady state for such a large temperature rise. Despite the thermal gradient, the Al-doped Ag films withstand thermal cycling at high temperatures (near their melting point) without initial damage. The abnormal temperature dip in the top-left corner of the GSST pattern is attributed to a change in the thermoreflectance coefficient, indicating that sublimation occurred during repeated aggressive thermal cycling. Furthermore, we measured the real-time heating and cooling response of our devices using 3.5 V - 50 $\mu$s pulses. The average response shown in Fig.~\ref{fig3}(e) leads to a cooling constant of 37.8 $\mu$s, which ultimately limits the microheater speed.

We last characterized the endurance of our Al-doped Ag microheater platform using the same TTI approach. In particular, we used 50~$\mu$s pulses at varying powers and duty cycle of  5\% to enable complete cooling.  As shown in Fig.~\ref{fig3}(f), after raising the temperature to nearly the melting point of GSST's ($\approx$900 K, as shown in Fig.~\ref{fig3}(d)), the failure occurs at about 107,650 cycles. At 600 K, the number of cycles increased dramatically, reaching 1.7 million before the microheater failed. For lower temperatures, near 400 K, we achieved more than 50 million cycles before stopping the experiment, with no failures or noticeable changes in the device. We also plot in Fig.~\ref{fig3}(d) the total accumulated heating time, defined as the total time during which the microheater holds the target temperature. In particular, we highlight that the microheater can operate for nearly 1 h without failure at temperatures around 400 K. Although our endurance experiment used electrical pulses, we expect a similar response under continuous operation. We show the devices before and after the cyclability test at different temperatures in Supplementary \ref{secS3}.\

\subsection{Nonvolatile switching of \texorpdfstring{Ge$_2$Sb$_2$Se$_4$Te}{GSST}: high-temperature stability}\label{subsec2_3}

\begin{figure}[!h]
\centering
\includegraphics[width=1.0\textwidth]{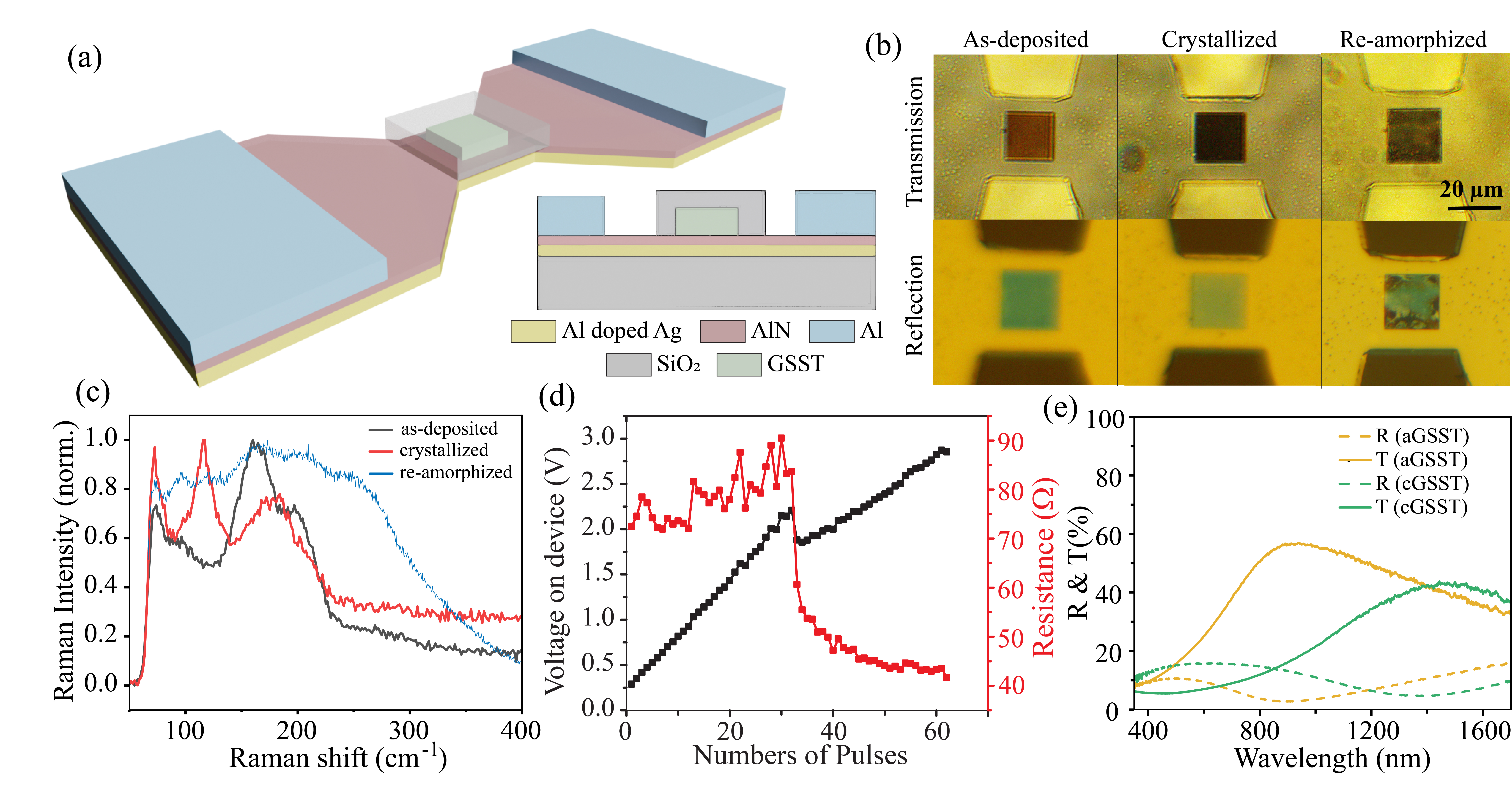}
\caption{\textbf{Reversible switching of GSST by 12 nm Al-doped Ag microheaters.} (a) The cross-section schematic of the device with GSST. (b) Transmission and reflection optical images of GSST reversible switching 30~$\times$~30~$\mu$m$^2$ under halogen illumination for better amorphous-crystalline contrast. (c) Raman spectra measured at the center of the PCM cell as-deposited (black), after a 2.2 V - 200 ms pulse (red), and after a 4.1V - 50~$\mu$s pulse (blue). (d) Applied voltage to the device and its resistance when applying a 50~$\mu$s-width amorphization pulse. (e) The Reflectance and transmittance spectra of GSST on 12 nm Al-doped Ag  in amorphous and crystalline states in the visible to NIR range}\label{fig2}
\end{figure}

We now shift focus to assess the optical performance of Al-doped Ag microheaters with embedded GSST cells in real device environments. The final device, considering metal contacts and protective capping, is sketched in Figure~\ref{fig2}(a) and imaged in Figure~\ref{fig2}(b). GSST exhibits slow crystallization kinetics \cite{aryana2021suppressed}, which is beneficial because, despite the microheater's relatively slow cooling rate, effective quenching is still possible and, thus, reamorphization. We demonstrate the reversible switching of GSST in Fig.~\ref{fig2}(b). Initially deposited in the amorphous state, the GSST square was transformed into a fully crystalline state by a 2.2 V, 200 ms-wide pulse---same pulse widely studied in Fig.~\ref{fig3}. We confirmed this transition by detecting the distinct Raman peak for crystalline GSST at 120 cm$^{-1}$, as shown in Fig.~\ref {fig2}(c). In real time, we monitored the device voltage and heater resistance during a voltage sweep using 50~$\mu$s pulses, as shown in Fig.~\ref{fig2}(d). A pronounced drop in both parameters was observed before the onset of crystallization, indicating that an annealing or conditioning pulse is necessary for reliable operation of the Al‑doped Ag heater. We hypothesize that this behavior results from silver grain growth at high temperatures, which reduces grain-boundary scattering. On the other hand, grain growth and dewetting can reduce the uniformity of the microheater, leading to failure during repeated heating cycles. \cite{kraft2001fatigue}.\

To return GSST to the amorphous state, we used a pulse of 4.1~V-50~$\mu$s to heat the material above 900 K and then quench. As observed in Fig.~\ref{fig3}(e), it takes approximately 12.5 $\mu$s for the device to cool from 900 K to 573 K (the crystallization temperature of GSST), which should suffice given the ms-scale crystallization kinetics of GSST \cite{huang2025optical}. The re-amorphized GSST was validated via Raman spectroscopy, as shown in Fig,~\ref{fig3}(c), where the 160~$cm^{-1}$ peak reappears after the 4.1~V-50~$\mu$s pulse. However, GSST is not entirely switched due to the non-uniform thermal profile. The corners of the GSST squares undergo higher temperatures than the center (see Fig.~\ref{fig2}(b)), resulting in damage \cite{popescu2025understanding}. This effect limits the switching performance; we achieved only 4 reversible cycles between the amorphous and crystalline states. To better analyze the evolution of the temperature profile within the GSST region, we performed time-dependent three-dimensional finite-element method (3D-FEM) simulations in COMSOL Multiphysics. The results are presented in the Supplementary ~\ref{secS2}. We found that a square geometry indeed leads to localized hotspots in the corners. Interestingly, we also demonstrate that a round GSST geometry could overcome the large temperature gradient and avoid damage, an approach we apply later to VO$_2$.

Further device engineering can improve most failure mechanisms in GSST-based devices \cite{popescu2025understanding}. Here, we instead focus on the performance of the microheater, which withstands aggressive melt-quenching thermal cycles at temperatures approaching 900 K. Moreover, as shown in Fig.~\ref{fig3}(f), the microheater endures nearly 10$^5$ cycles, which is one order of magnitude larger than the 10$^4$ reported for metasurfaces based on GSST \cite{popescu2025understanding}. 
Lastly, in Fig.~\ref{fig2}(e), we show the nonvolatile optical modulation of this device in the visible and near-infrared ranges, including the first demonstration of tunable transmission. In the amorphous state, a maximum of 57\% transmission is obtained at 940 nm. Once the GSST is switched to the crystalline state, the transmittance peak shifts to 1500 nm, accompanied by a drop in overall transmittance of 70\% when averaged over the 580-920 nm wavelength range.

\subsection{Volatile switching of \texorpdfstring{VO$_2$}{VO2}: low-temperature endurance}\label{subsec2_4}
\begin{figure}[!h]
\centering
\includegraphics[width=0.95\textwidth]{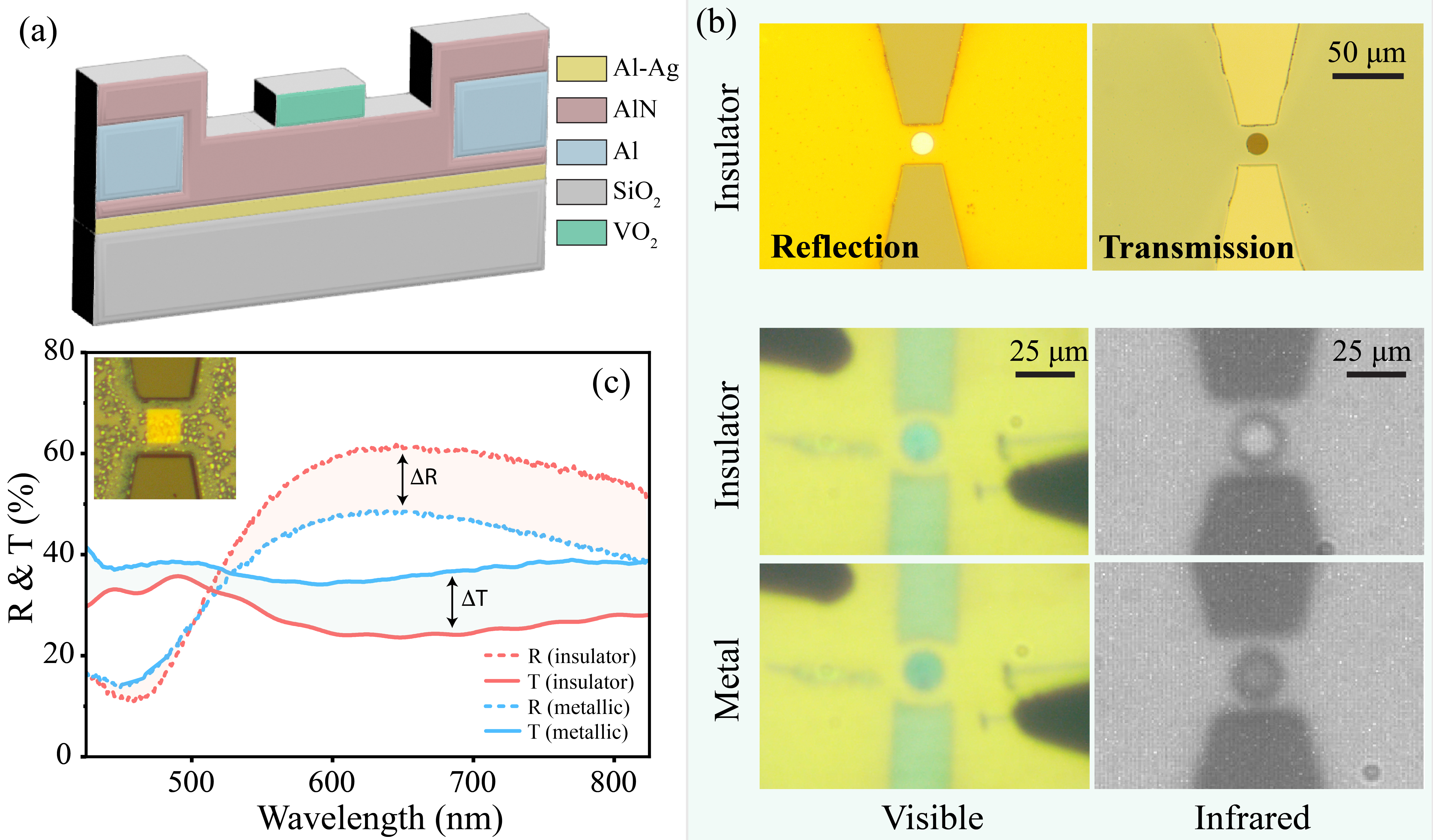}
\caption{\textbf{Reversible switching of VO$_2$ by 12~nm Al-doped Ag microheaters.} Reversible switching of VO$_2$ by 12~nm Al-doped Ag microheaters.(a) The cross-section schematic of the Al-doped Ag microheaters with embedded VO$_2$.(b) Transmission and reflection optical images of 30 nm-thick VO$_2$ at room temperature, and optical images in reflection before (insulator) and after heating (metallic), in the visible and infrared ranges. (c) Transmittance (solid line) and reflectance (dashed line) of the 30 nm VO$_2$ at insulator phase and spectra measured at the center of the VO$_2$ cell at insulator (red) and metallic (blue) phases. The inset optical image shows the device in which this data was collected.}
\label{fig4}
\end{figure}

\texorpdfstring{VO$_2$}{VO2} is a volatile phase-change material with low phase transition temperature (approximately 65 $^\circ$C) and large refractive index change via Insulator-to-Metallic transition (IMT) \cite{king2024electrically,wei2024tunable}. Given that such properties are ideal for our Al-doped Ag platform, especially given the requirement of a temperature increase under 100 K, we integrated VO$_2$ into Al-doped Ag heaters, using the device configuration sketched in Fig.~\ref {fig4}(a).  Since VO$_2$ is deposited by annealing VO$_x$ in an Ar/O$_2$ at high temperature (450 $^\circ$C), the AlN protective layer becomes crucial. In particular, we found that AlN thicknesses below 50 nm resulted in damage to the Al-doped Ag film---see Supplementary Section ~\ref{secS1}. Therefore, we chose to first coat the Al-doped Ag film with a thin 5 nm-thick AlN, pattern the Al contacts, and then deposit an extra 60~nm of AlN, leaving the VO$_2$ deposition and annealing as the last step in the fabrication process. The VO$_2$, AlN, and Al-doped Ag film stack was additionally designed to form a Fabry-Perot cavity, enhancing the optical response in the visible spectrum.\

In Fig ~\ref{fig4}(b), we show an optical image under halogen illumination for optimal contrast (there is no significant difference in optical properties and response for wavelengths below 550 nm) for both transmittance and reflectance.  Moreover, in Fig.~\ref{fig4}(b), we show the reflectance during VO$_2$ reversible switching for two scenarios: the visible spectrum using a yellow filter centered close to the maximum change around 600 nm, and the infrared range. Supplementary video 1 shows the device in operation under visible light illumination. For reversible switching, we applied pulses of 1.2 to 2.6 V to microheaters with areas of 25~$\times$~25~$\mu$m$^2$ to 100~$\times$~100~$\mu$m$^2$, achieving reconfigurable IMT, demonstrated but not limited to a frequency of 25 Hz. We performed reversible switching for 10 min at 25 Hz, for a total of 4000 thermal cycles with no change in device performance---as predicted in the endurance test in Fig.~\ref{fig3}(f).\

Furthermore, in Fig.~\ref{fig4}(c), we show the transmission and reflection spectra measured in the center of the VO$_2$ pattern. The maximum transmission is at 480~nm for both VO$_2$ states, while the largest transmission contrast of 30\% is observed in the 550 - 825 nm range. We also investigated this spectral response via Lumerical FDTD simulations in Supplementary ~\ref{secS4}. We found that our experimental data fit both transmittance and reflectance in the 425~nm to 600~nm range, with a blue-shift, while the spectra in longer wavelength ranges are flatter than the simulated data. The differences in intensities and the spectrum for the longer-wavelength region($> 600~nm$) are attributed to differences between the simulated and experimental refractive index for VO$_2$ and the surface roughness---see Supplementary ~\ref{secS1}.

\subsection{State-of-the-art of microheaters for tunable materials}\label{subsec2_5}

To provide a comparison and contextualize our results, we summarize in Table.~\ref{tab1} the most important achievements in microheaters for free-space devices based on phase-change materials.  Owing to the high conductivity of silver, the 12~nm Al-doped Ag heater in this work achieves significantly lower switching voltages compared with W and Ti/Pt heaters with a comparable footprint. Relative to doped-Si and graphene heaters, which often require higher voltages or exhibit high sheet resistance, the present design provides moderate resistance (40–50 $\Omega$) and therefore efficient Joule heating. Transparent conductive oxides generally rely on much thicker films (200–450 nm), which can increase optical loss, reduce tuning speed due to higher thermal mass, and introduce parasitic interferences and other effects. In contrast, ultra-thin Al-doped Ag films are a low-perturbative alternative that minimizes thermal mass while maintaining low-voltage operation. For VO$_2$, comparable transition voltages have been reported (2.6 V); however, Al-doped Ag achieves this with substantially reduced thickness, underscoring its suitability for compact, tunable devices. Moreover, our approach, combining Al-doped Ag microheaters with VO$_2$, represents the first tunable device based on this PCM operating in transmission at visible wavelengths, due to the dual role of the Al-doped Ag film as both a transmissive microheater and a partial reflector in a Fabry-Perot cavity. 

\begin{table}[htbp]
\centering
\footnotesize
\caption{Comparison of microheaters for PCMs in free-space optics}\label{tab1}%
\begin{tabular}{@{}llllllll@{}}
\toprule
Reference & Material &Thick (nm)& R ($\Omega$) & Size ($\mu$m$^2$)& PCM & Am. Pulse & Cry. Pulse\\
\midrule
\cite{qu2024effect}  & W & 100 & 6.6-9.2 & 20$\times$20 & Ge$_2$Sb$_2$Te$_5$ & 20V-200~ns  & 11~V-3$\mu$s \\
\cite{abdollahramezani2022electrically}\footnotemark[1]  & W & 50  & 6.25   & 10.5$\times$10.5 & Ge$_2$Sb$_2$Te$_5$ & 3.4~V-200ns  & 1.7~V-200$\mu$s \\
\cite{zhang2019broadband}    & Ti/Pt & 50/20 & N/A & 10$\times$10& Ge$_2$Sb$_2$Se$_4$Te & 24~V-1$\mu$s  & 13~V-500$\mu$s \\
\cite{zhang2021electrically}  & Ti/Pt & 50/20 & 20 & 200$\times$200& Ge$_2$Sb$_2$Se$_4$Te & 20~V-5$\mu$s  & 10~V-500~ms \\
\cite{popescu20252daddressablemidinfraredmetasurface}    & Doped Si & 220 & N/A & 120$\times$120& Ge$_2$Sb$_2$Se$_4$Te  & 25.5~V-13$\mu$s  & 18~V-11$\mu$s \\
\cite{tara2025electricallyreconfigurablenonvolatiletransmissive}   & Doped Si & 55  & 250 & 20$\times$20 & Sb$_2$S$_3$  & 25~V-9$\mu$s  & 12.5~V-100~ms \\
\cite{youngblood2021reconfigurable}\footnotemark[1] & FTO & 450 & 23.1 & 100$\times$ 100 & Sn-doped Ge$_{20}$Te$_{80}$  & 5.3~V-70$\mu$s  & 4.65~V-130$\mu$s \\
\cite{xu2026nonvolatile} & In$_2$O$_3$ & 200 & N/A & 40$\times$ 40 & Sb$_{2}$Se$_{3}$  & 4.5~V-20$\mu$s  & 2~V-600$\mu$s \\
\cite{rios2021multi}    & Graphene & 0.34 & 2.5~k & 10$\times$10 & Ge$_2$Sb$_2$Se$_4$Te  & 7.5~V-13$\mu$s  & 6~V-20~ms \\
this work    & Al-doped Ag & 12 & 40-50 & 35$\times$35& Ge$_2$Sb$_2$Se$_4$Te  & 4.1~V-50$\mu$s  & 2.2~V-200~ms \\
\cite{zangeneh2023electrically}   & ITO & 380 & N/A & 100$\times$200 & VO$_2$ & N/A  & 3.6V \\
\cite{guo2024durable}   & ITO & 50 & 400 & 155$\times$120 & VO$_2$ & N/A  & 1.8V \\
this work    & Al-doped Ag & 12 & 80-120 & 100$\times$100 & VO$_2$ & N/A  & 2.6V \\
\botrule
\end{tabular}
\footnotemark[1]{Data in the row is calculated based on the sheet resistance or the resistivity provided in the reference.}
\end{table}

\section{Conclusions}\label{sec3}
We demonstrated that Al-doped Ag films as thin as 7~nm serve as effective ultra-thin film metal heaters for free-space optics, displaying smooth surface, outstanding thermal stability, easy and high-yield fabrication, $\sim$80\% transmission in the visible spectrum, and low resistance (e.g., 40 $\Omega$ in 35~$\times$~35~$\mu$m$^2$ microheaters). We verified thermal stability up to 10$^5$ to 10$^7$ cycles at temperatures nearing $\sim$900 K (with failure) and $\sim$400 K (without failure), respectively. Furthermore, we integrated two phase-change materials, GSST and VO$_2$, into Al-doped Ag microheaters to enable nonvolatile and volatile spectral modulation via temperature-triggered phase transitions. We demonstrated reversible switching of GSST, encompassing melting over 900 K and quenching with a maximum transmittance of 57\% at 940 nm and a 40\% amplitude difference between the two states. Better thermal performance was demonstrated at lower temperatures, with multiple cycles of VO$_2$’s metal-to-insulator transitions at around 65$^\circ$C. To do so, we used a Fabry-Perot-type device that represents the first demonstration of VO$_2$-driven optical transmission modulation in the visible spectrum. \ 

Our approach already demonstrates outstanding performance compared to similar microheater platforms. Table 1 compares several platforms and highlights the advantages of Al-doped Ag for low-voltage operation in large-area microheaters, making it the thinnest microheater yet demonstrated, which enables low-perturbative integration into multi-layer or metasurface devices without sacrificing thermal and electrical conductivity. In addition, considering the demonstrated ductility of ultra-thin silver films \cite{ma2024pushing}, we anticipate Al-doped Ag films and devices to also be compatible with flexible substrates.  We note that further work could significantly improve the endurance of Al-doped Ag films during high-temperature operation by engineering surface diffusion, grain growth, and dewetting near to their melting point. This work introduced a new class of microheaters that could be utilized in a wide range of nanophotonic devices, including optical filters and tunable metasurfaces for spectral, spatial, and polarization control. 
\
\
\section{Methods}\label{sec4}
\subsection{Device fabrication}\label{subsec4_1}
We fabricated the Al-doped Ag heater using a standard lift-off process combining photolithography and magnetron sputtering techniques. Briefly, 1 cm $\times$ 1 cm fused silica pieces were cleaned with acetone followed by isopropyl alcohol. For multilayer fabrication, gold alignment markers (cross patterns) were first patterned. A photoresist SPR220 3.0 was spin-coated at 3000~rpm and soft-baked at 115$^\circ$C for 90 seconds. The photoresist was then exposed using a direct laser writer equipped with a 365 nm laser (Heidelberg MLA150). Following a post-exposure bake at 115$^\circ$C for 90 seconds, the pattern was developed in AZ300 developer for 1 minute.
Subsequently, 10-nm chromium and 50-nm gold were deposited via e-beam evaporation, and lift-off was performed in acetone. For Al-doped Ag patterning, the same lift-off procedure was followed. The Al-doped Ag layer was deposited using a magnetron sputtering system (Denton Vacuum Discovery 550 $\&$ AJA 1800 Sputtering unit) as described in \cite{zhang2017high}.  Aluminum and silver were simultaneously sputtered at rates of 0.08~nm/s and 1.1~nm/s, respectively. The pattern was then revealed by lift-off in acetone. 
A 5~nm AlN layer is sputtered to prevent the Al-doped Ag heater from being damaged during PCM patterning. A 120~nm-Al electrode is used as the contact on top of the heater. The patterning resist S1813 was spin-coated at 4000~rpm with 3 min pre-bake at 180$^\circ$C and 1-min postbake at 100$^\circ$C. The exposure is accomplished by a 375~nm laser (Heidelberg MLA150), followed by 45s developing in CD-26 developer. The 25-nm-thick Ge$_2$Sb$_2$Se$_4$Te thin-film was deposited onto the microheaters using an AJA Orion-3 Ultra High Vacuum Sputtering system at room temperature. The VO$_2$ was synthesized by reactive sputtering vanadium and post-annealing in an O$_2$ atmosphere as described in \cite{hallman2021sub}. The PCM cell patterning was performed using electron beam lithography on an Elionix ELS-G100 system with a Ma-N 2403 negative resist, followed by reactive-ion etching with CF$_4$.  For GSST devices, a 60nm-thick SiO$_2$ is capped as a passivation layer on the top.\  

\subsection{Electrical characterization:}\label{subsec4_2} The electrical sheet resistance was measured with a 4-point probe measurement using a Signatone Pro4 Resistivity Test System, and device resistance was monitored using a Keithley 2400. The electrical switching of the PCM was achieved using a probe station with a Moku:Pro (Liquid Instruments) as the pulse generator. To overcome the maximum voltage limitations of the Moku:Pro, the output line was connected to a wide-bandwidth current feedback amplifier (THS3491DDA, Texas Instruments) to amplify arbitrarily pulses. A Keithley 2400 and a Rockseed RS305D were used as DC power sources.\ 

\subsection{Optical characterization:}\label{subsec4_3} The spectral measurements were captured in a custom-built optical setup coupled to a F40-UVX Filmetrics system (KLA). A Deuterium-tungsten light source (LS-DT2, KLA) was focused onto the device using a 15$\times$ microscope (ME520TC-18M3, Amscope) for reflectance measurements and a coaxial vision system (Zoom 6000, Navitar) with up to 30$\times$ magnification for transmittance measurements. To simultaneously visualize the device and acquire the reflected signal, we used a beam splitter to evenly split the light between a CCD camera (Amscope MU500-HS) for imaging and a Filmetric F40-UVX spectrometer for spectral acquisition. The measured reflection spectra were normalized by dividing the collected signal by the intensity of a reference beam reflected from a standard Si wafer with the same illumination spot size.\ 

We measured the complex refractive index using a J.A. Woollam M-2000D Spectroscopic Ellipsometer and performed a Tauc–Lorentz fitting from the general oscillator (Gen-Osc) model. \ 

\subsection{Transient thermoreflectance Imaging (TTI):}\label{subsec4_4}TTI was used to characterize the microheater’s transient thermal dynamics\cite{sun2025microheater}. A single-wavelength LED (530~nm) was pulsed to measure the surface reflectance change of the microheater. A layer of silver paste is applied on the back surface of the chip to enhance the reflectance from the sample chip. The accuracy of TTI depends on determining the thermoreflectance coefficient $C_{\mathrm{TH}}$, assuming a linear relationship between temperature rise and reflectance change. The change in thermoreflectance ($\Delta$R/R) was measured for a given set of temperature rises by increasing the stage temperature from 20~$^{\circ}$C to 120~$^{\circ}$C. The surface temperature rise is independently measured using a thermocouple positioned near the device. Using an iterative approach, the C$_{TH}$ of the heater region was monitored, and calibrations were repeated until the C$_{TH}$ value converged. Averaging over multiple pixels, the standard deviation, with 95{\%} confidence intervals, was used to estimate the C$_{TH}$ uncertainty. For the 530~nm excitation, the C$_{TH}$ of the Al-doped Ag microheater with both phases of GSST was measured to be $-1.12\times10^{-4}~K^{-1}$ and $-5.46\times10^{-4}~K^{-1}$, respectively. A pulsed IV system was used to electrically bias the microheater for TTI. Based on a lock-in approach, an internal trigger was used to synchronize the pulsed IV with the TTI system. A 5\% duty cycle was implemented to ensure that the heater has sufficient time to cool down.\

\subsection{Raman characterization}\label{subsec4_5}
The PCM phase identification was performed using surface-enhanced Raman spectroscopy (Yvon Jobin LabRam ARAMIS) with a 532 nm laser, a 2400 lines/mm grating, and a 100× long-working-distance objective.\

\subsection*{Acknowledgements}

This work has been supported by ARO (84471-PE-II), AFOSR (FA9550-24-1-0064), the National Science Foundation (ECCS-2210168/2210169, ECCS-2132929, and DMR-2329087/2329088), and is supported in part by funds from federal agencies and industry partners as specified in the NSF Future of Semiconductors (FuSe) program. C.R.O. acknowledges support from the Minta Martin Foundation through the University of Maryland. S.W.M and C.S.W. deposited VO$_2$ in the Vanderbilt Institute of Nanoscale Science and Engineering.

\renewcommand{\thefootnote}{\fnsymbol{footnote}}
\newpage

\bibliography{sn-bibliography}

\newpage

\setcounter{section}{0}  
\renewcommand{\thesection}{S\arabic{section}}
\renewcommand{\theHsection}{S\arabic{section}}

\setcounter{figure}{0}  
\renewcommand{\thefigure}{S\arabic{figure}}
\renewcommand{\theHfigure}{S\arabic{figure}}

\setcounter{table}{0}  
\renewcommand{\thetable}{S\arabic{table}}
\renewcommand{\theHtable}{S\arabic{table}}

\setcounter{equation}{0}  
\renewcommand{\theequation}{S\arabic{equation}}
\renewcommand{\theHequation}{S\arabic{equation}}
\begin{center}
    {\Large \textbf{Supplementary Information}}
\end{center}

\section{Temperature profile simulation }\label{secS2}

\begin{figure}[!h]
\centering
\includegraphics[width=1\textwidth]{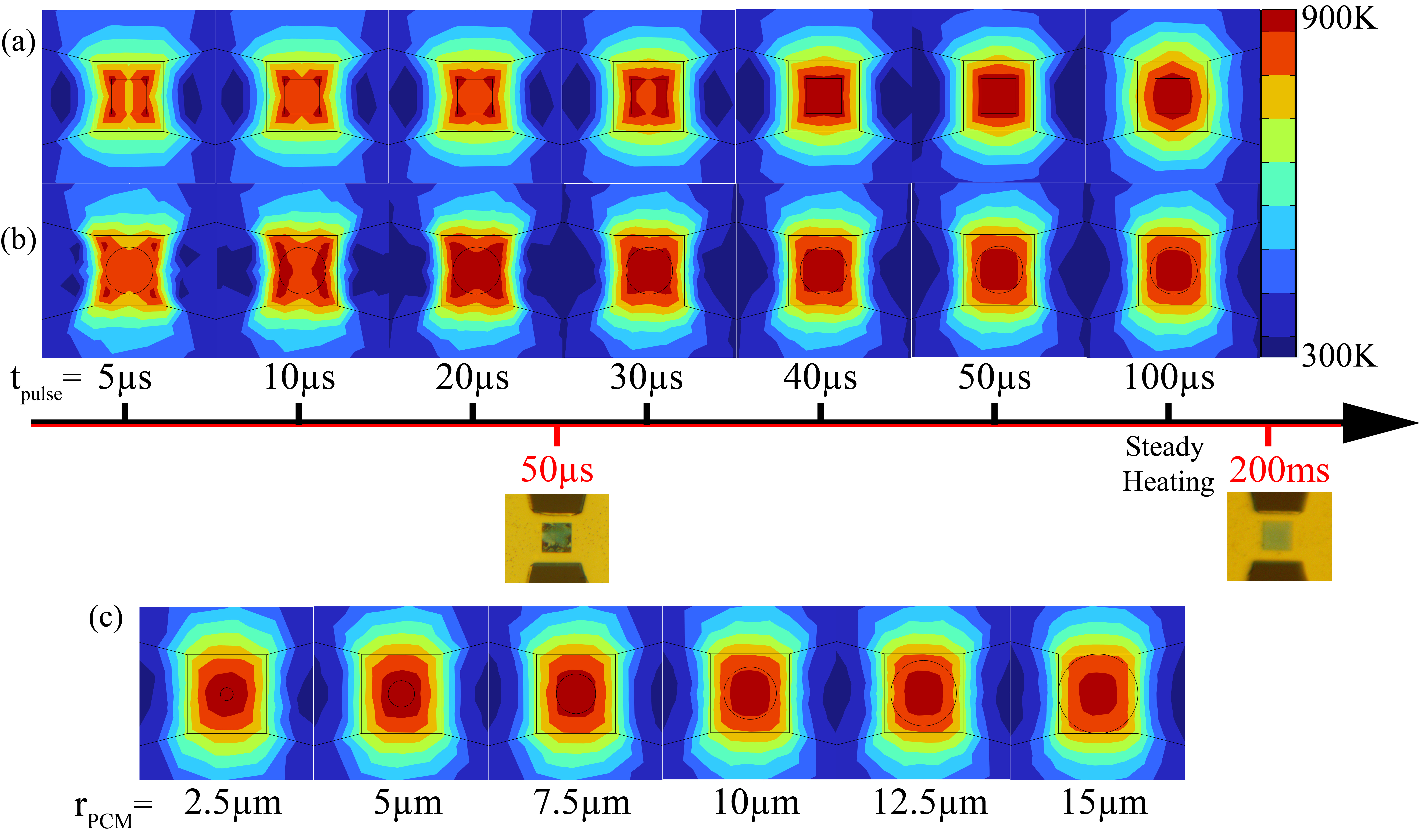}
\caption{\textbf{Simulated temperature profile applied on Al-doped Ag heater.} The temperature profile on the top surface of a 30~$\times$~30~$\mu$m$^2$ Al-doped Ag heater with the GSST pattern on the top. t$_{Heater}$=12~nm, t$_{Electrode}$=80~nm, t$_{AlN}$=5~nm, t$_{SiO_{2}capping}$=50~nm. (a) Square GSST pattern with different pulse widths, W$_{GSST}$=15~$\mu$m. (b) Round GSST pattern with different pulse widths, R$_{GSST}$=7.5~$\mu$m. (c) Round pattern with different radius.}\label{figS2}

\end{figure}

To simulate the Joule heating distribution across the phase-change material (PCM) region, we performed time-dependent three-dimensional finite-element-method (3D-FEM) simulations using COMSOL Multiphysics. Figure~\ref{figS2} illustrates the surface temperature distribution of an Al-doped Ag microheater integrated with a 25~nm-thick GSST pattern. To optimize the cyclability of the PCM and the performance of the microheater, the temperature distribution at the end of the electrical pulse must be not only spatially uniform across the PCM region but also established as rapidly as possible.

Figures~\ref{figS2}(a) and (b) show the temperature profiles resulting from electrical pulses with different widths, where the applied voltages are adjusted to achieve comparable peak temperatures. The simulations indicate that devices with a circular PCM pattern reach a more uniform temperature distribution before the thermal profile reaches steady state, making this geometry more suitable for our application. In addition, the lateral dimensions of the PCM play a critical role. As shown in Fig.~\ref{figS2}(c), to obtain a uniform temperature distribution at the end of a 100-$\mu$s pulse, the radius of the GSST pattern should be less than 60~\% of the width of the square microheater.

\section{Optical images for Al-doped Ag heater in cyclability characterization}\label{secS3}

\begin{figure}[!h]
\vspace{-8pt}
\centering
\includegraphics[width=1\textwidth]{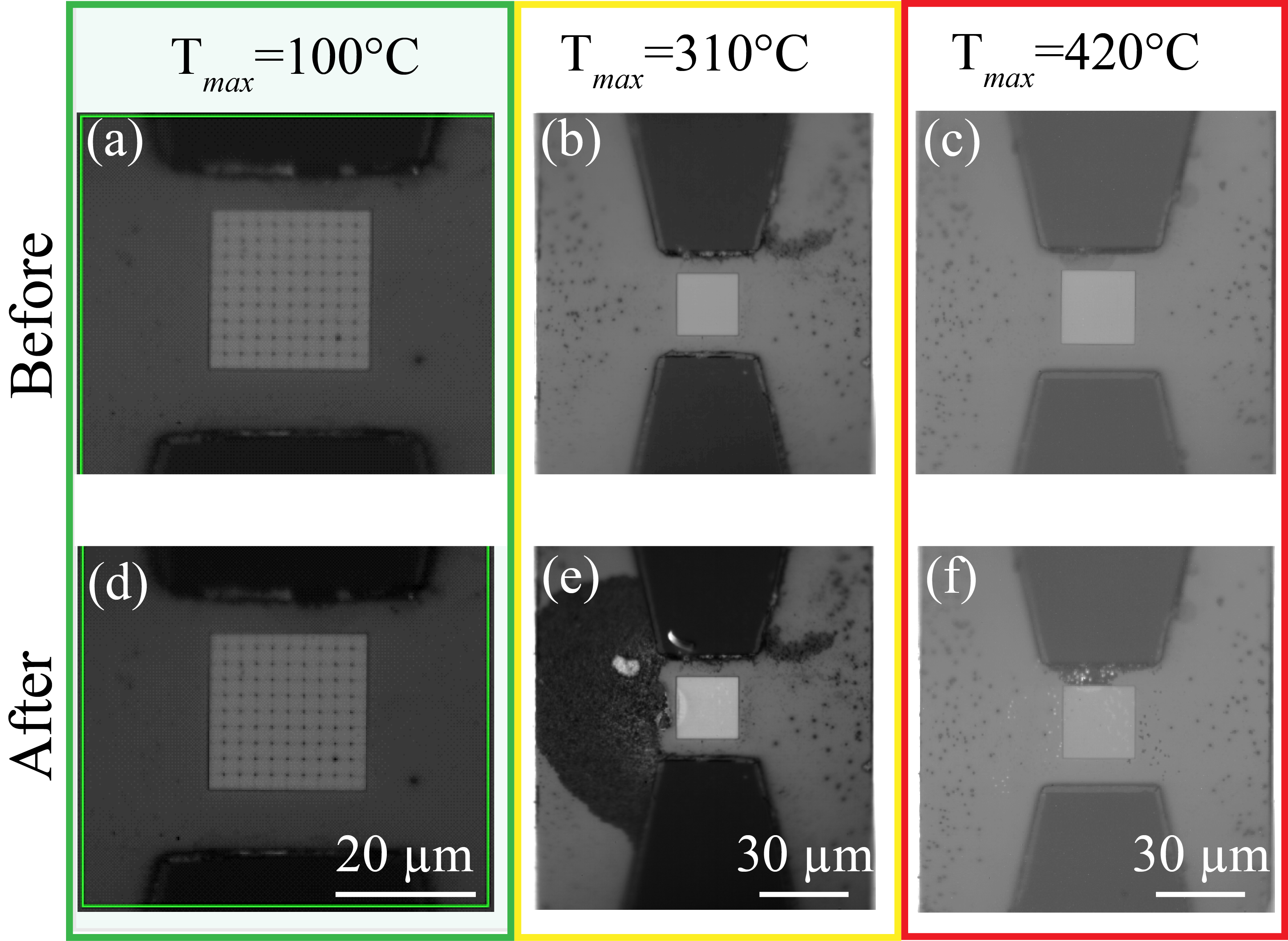}
\caption{\textbf{Optical images of before and after cyclability test of Al-doped Ag devices with GSST squares}. (a) T$_{max}$=100 $^{\circ}C$. (b) T$_{max}$=310 $^{\circ}C$. (c) T$_{max}$=420 $^{\circ}C$.} \label{figS3}
\end{figure}

Figure~\ref{figS3} presents optical micrographs of representative microheaters before (a--c) and after (d--f) cyclability tests under different maximum temperature conditions. For $T_{\max} = 100^\circ\mathrm{C}$ in Fig~\ref{figS3}. (a) \& (d), the device morphology remains essentially unchanged after cycling, with no observable delamination, cracking, or deformation, indicating excellent structural stability at low thermal load. At $T_{\max} = 310^\circ\mathrm{C}$ in Fig~\ref{figS3}. (b) \& (e), noticeable material accumulation and surface modification appear near the electrode edges after cycling, suggesting thermally induced diffusion, oxidation, or partial material reflow under repeated Joule heating. When the maximum temperature is increased to $T_{\max} = 420^\circ\mathrm{C}$ in Fig~\ref{figS3}. (c) \& (f), more pronounced morphological changes are observed, including edge roughening and localized surface degradation in the active region, consistent with intensified thermal stress and possible material phase instability. The progressive structural evolution with increasing $T_{\max}$ highlights the temperature-dependent reliability limits of the microheater and underscores the importance of thermal budget optimization for long-term device endurance.

\section{Scanning Electron Microscope(SEM)}\label{secS1}
\begin{figure}[!h]
\centering
\includegraphics[width=0.75\textwidth]{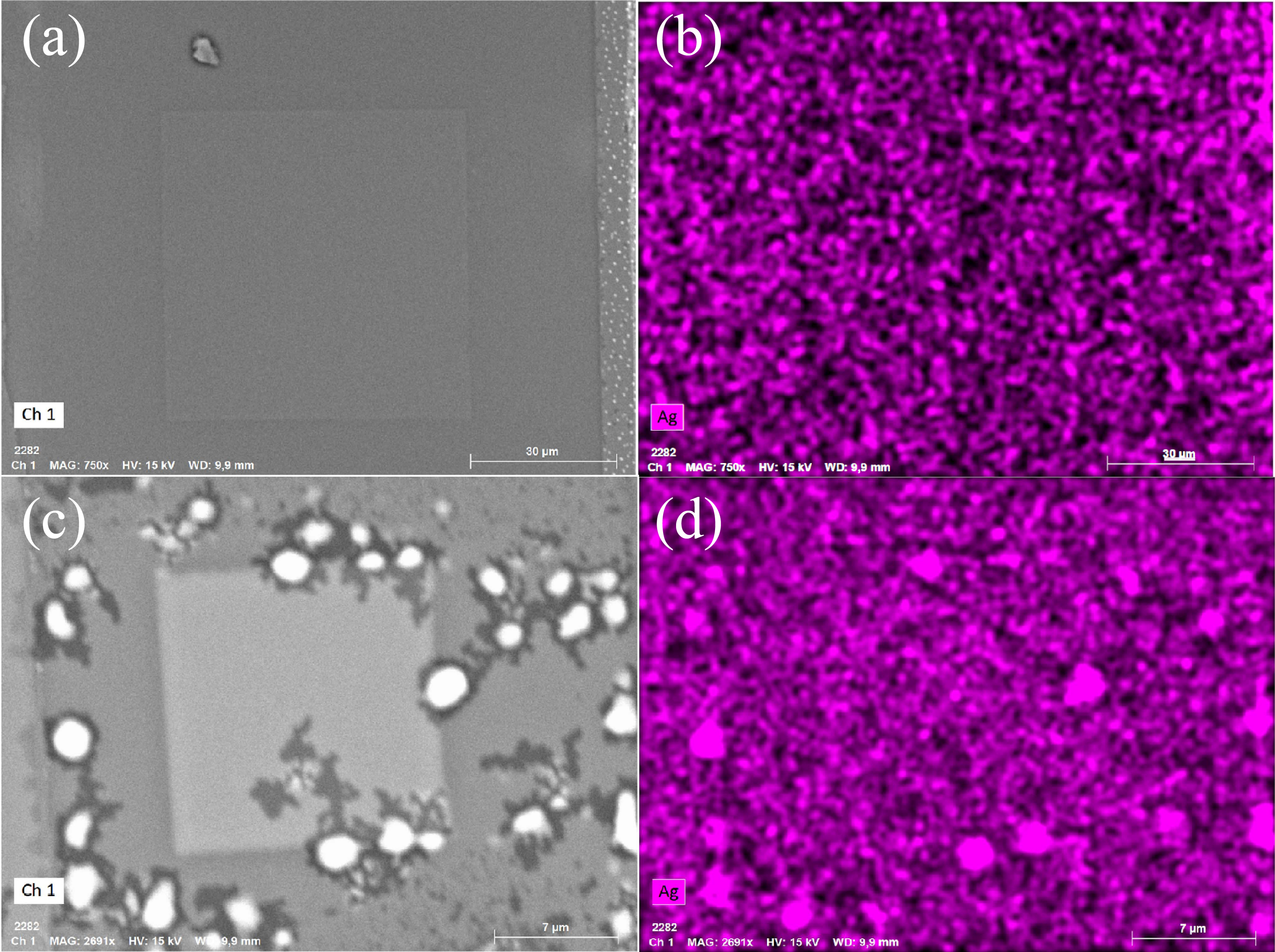}
\caption{\textbf{Scanning Electron Microscope(SEM)-Energy Dispersive X-ray Spectroscopy(EDS) analysis of Al-doped Ag heater's failure during VO$_2$'s deposition.} (a) The SEM image of a 100~$\times$~100~$\mu$m$^2$ Al-doped Ag heater. (b) The EDS elemental mapping of Ag of the microheater, which is shown in (a). (c) The SEM image of a damaged 25~$\times$~25~$\mu$m$^2$ heater after VO$_2$'s deposition, while the EDS elemental mapping of silver of the broken microheater is shown in (d). } 
\label{figS1}
\end{figure}

Fig.~\ref{figS1}(a) presents scanning electron microscope (SEM) images of both intact and damaged Al-doped Ag microheaters coated with a 70~nm-thick VO$_2$ film. A smooth and uniform surface is observed for the intact Al-doped Ag microheater and the VO$_2$ layer, as shown in Fig.~\ref{figS1}(a). In contrast, a fractured microheater is visible in Fig.~\ref{figS1}(c), which is attributed to the high-temperature annealing process used for VO$_2$ crystallization. 

To investigate the large clusters that formed after fabrication, energy-dispersive X-ray spectroscopy (EDS) analyses were performed on both devices, as shown in Fig.~\ref{figS1}(b) and (d). The EDS results reveal the formation of Ag-rich particles, leading to increased surface roughness and nonuniform morphology. Similar phenomena have been reported in previous studies \cite{kulczyk2014investigation, ye2019plasmonics}, where silver diffusion into the AlN layer or along the interfaces of the multilayer structure was identified as the underlying mechanism. To suppress Ag diffusion, replacing the AlN barrier layer with TiN represents a promising approach due to its superior diffusion-blocking capability.

\section{Optical Spectrum between FDTD simulation and Experimental results} \label{secS4}

\begin{figure}[!h]
\vspace{-8pt}
\centering
\includegraphics[width=0.75\textwidth]{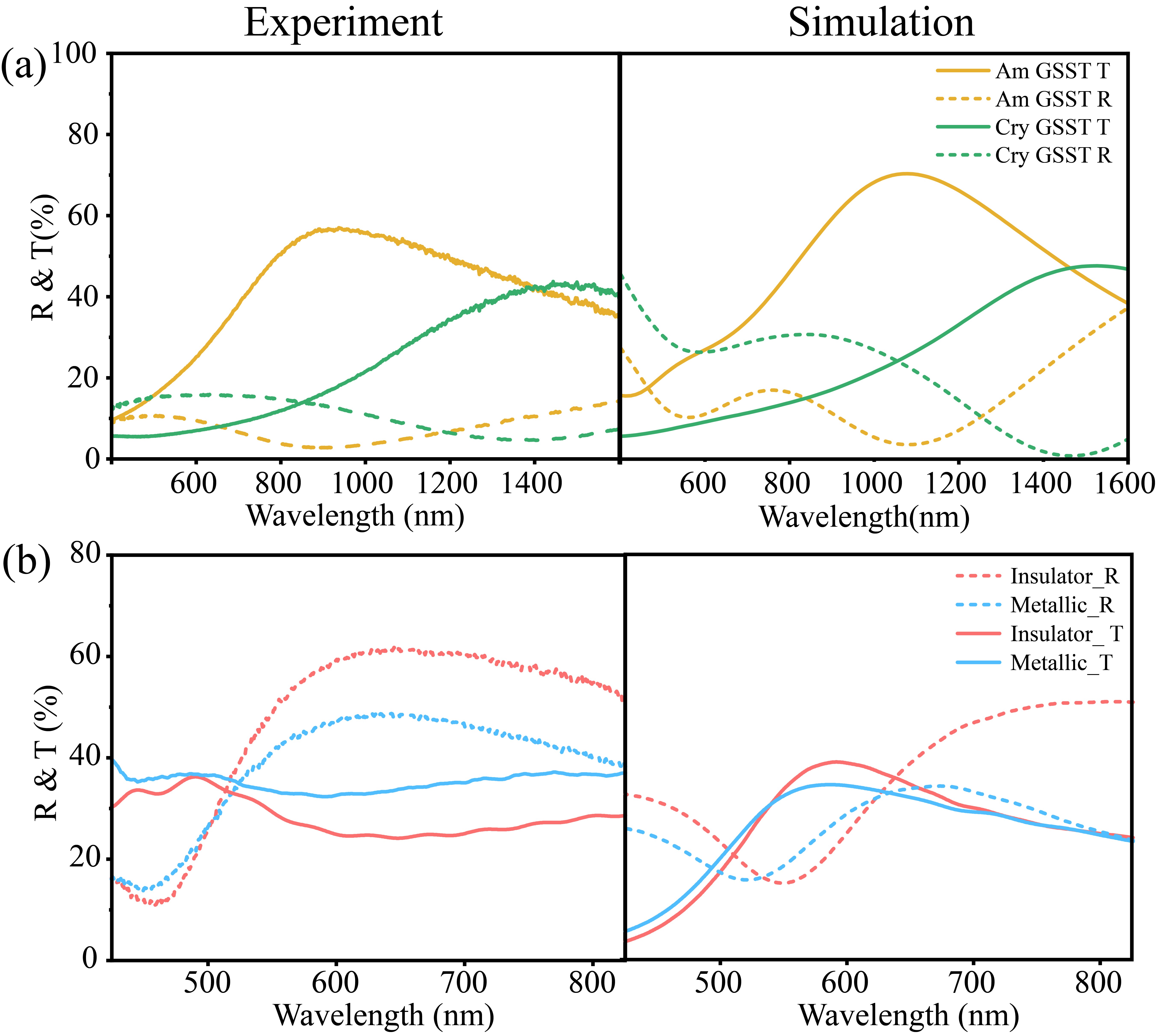}
\caption{\textbf{Simulated reflectance and transmittance spectrum of Al-doped Ag devices with PCMs}. (a) Al-doped Ag with GSST (b) Al-doped Ag with VO$_2$. } \label{figS4}
\end{figure}
 Lumerical FDTD simulations were conducted to study the optical response of the Al-doped Ag heater with PCMs. As shown in the Fig.\ref{figS4}, the left panels present the experimental results of reflectance (R) and transmittance (T), while the right panels show the corresponding numerical simulations. Fig.\ref{figS4}(a) corresponds to GSST in its amorphous (Am) and crystalline (Cry) states, and the bottom row corresponds to VO$_2$ in its insulating and metallic phases. 

For devices with GSST, the experimental results indicate that the amorphous state exhibits relatively high transmittance in the near-infrared region (approximately 800–1200 nm), reaching a maximum of 55{\%} near 900~nm, while the reflectance remains less than 15\%. In contrast, the crystalline state shows reduced transmittance and enhanced reflectance, particularly at wavelengths above 1000~nm, where the reflectance gradually increases to over ~40\%. The simulations reproduce these overall trends well: the amorphous state maintains significantly higher transmittance than the crystalline state in the mid-to-near-infrared region, and the crystalline-state reflectance increases toward longer wavelengths. The relative spectral contrast and phase-dependent modulation are in good agreement with the experimental observations.

For devices with VO$_2$, the experimental spectra show that in the insulating phase the device exhibits relatively high reflectance (50–60\%) from 540~nm to 825~nm, while the transmittance peaks at 480~nm (35\%). Upon transition to the metallic phase, a quasi-flat transmittance is observed from 450~nm to 850~nm, and the reflectance spectrum is reshaped, consistent with the free-carrier response characteristic of the metallic state. The simulations reproduce this phase-dependent spectral redistribution: the insulating phase maintains higher reflectance in the red region, while the transmittance remains relatively uniform across most of the visible spectrum. Compared with the simulated data, the experimental reflectance spectra exhibit a slight blue shift in peak position and minor variations in magnitude, and a gradual decay trend after the peak wavelengths. These discrepancies may arise from surface roughness and grain-size distribution, as discussed in Supplementary~\ref{secS1}.

\end{document}